\documentclass{article}

\usepackage[preprint]{neurips_2022}

\usepackage{latexsym}

\usepackage{mathtools} 
\usepackage{booktabs} 
\usepackage{tikz} 

\usepackage{float} 

\usepackage{xcolor}
\usepackage[normalem]{ulem}

\usepackage{times}

\usepackage{subfigure}

\usepackage{sidecap}

\usepackage{hyperref}

\usepackage[most]{tcolorbox}
\usepackage{enumitem}

\definecolor{kpback}{rgb}{0.99,0.99,0.985} 
\definecolor{kpframe}{rgb}{0.82,0.82,0.82} 
\definecolor{kptitle}{rgb}{0.20,0.20,0.20} 

\newtcolorbox{keypointbox}[1][]{%
  enhanced,
  breakable,
  colback=kpback,
  colframe=kpframe,
  coltitle=kptitle,
  boxrule=0.35pt,              
  arc=0.5pt,                   
  outer arc=0.5pt,
  boxsep=4pt,                  
  left=10pt,
  right=10pt,
  top=7pt,
  bottom=7pt,
  title={#1},
  fonttitle=\bfseries\scshape\small,
  before skip=10pt plus 2pt minus 2pt,
  after skip=10pt plus 2pt minus 2pt,
  before upper={%
    \fontsize{9.2}{10.5}\selectfont
    \itshape
    \setlength{\parindent}{0pt}%
    \setlength{\parskip}{3pt}%
    \setlist[itemize]{leftmargin=1.4em, topsep=2pt, itemsep=2pt, parsep=0pt}%
  },
}

\title{
The Road of Adaptive AI for \\ Precision in Cybersecurity
}

\author{
  Sahil Garg\\
  AI Research at Averlon\\
  Redmond, WA\\
  \texttt{sahil.garg@averlon.io, sahil.garg.cs@gmail.com}\\
}

\usepackage{xcolor}
\usepackage{ulem} 
\usepackage{natbib}

\begin{document}

\maketitle

\begin{abstract}
Cybersecurity's evolving complexity presents unique challenges and opportunities for AI research and practice. This paper shares key lessons and insights from designing, building, and operating production-grade GenAI pipelines in cybersecurity, with a focus on the continual adaptation required to keep pace with ever-shifting knowledge bases, tooling, and threats. 
Our goal is to provide an actionable perspective for AI practitioners and industry stakeholders navigating the frontier of GenAI for cybersecurity, with particular attention to how different adaptation mechanisms complement each other in end-to-end systems.
We present practical guidance derived from real-world deployments, propose best practices for leveraging retrieval- and model-level adaptation, and highlight open research directions for making GenAI more robust, precise, and auditable in cyber defense.
\end{abstract}
    
\paragraph{Disclaimer:} The ideas and analysis presented here are subjective. We share them based on our experience of establishing robust and efficient pipelines of generative AI for cybersecurity. In light of the age of generative AI, the objective of this document is not to provide generic descriptions of GenAI techniques, but rather to explain their practical relevance for specific contexts, and to illustrate where particular choices have worked well or poorly in our own deployments.
                    
\section{Introduction}
\label{sec:intro}

It has been an adventurous journey in the wilderness of cybersecurity, in the continual pursuit of my passion for problem-driven AI research. Although it may seem otherwise at first glance to newcomers to the domain, cybersecurity is highly rich in its complexities, with enough challenging AI problems to engage AI researchers for a lifetime~\citep{coelho2024survey, xu2024llmcyber, ferrag2024llm_cyber_survey}. 

In the past decade, one of the most challenging and complex domains where AI has played a significant role has been healthcare. Similarly, the cybersecurity domain is not only complex but also dynamic and constantly evolving, which makes it particularly interesting from an AI research perspective. While we aspire to a future of dramatically reduced disease burden, driven by focused AI research efforts in medicine, cybersecurity is fundamentally a never-ending game. Unlike in the field of medicine, the problems we tackle today can become obsolete tomorrow. This continual obsolescence is not only exciting but also an important consideration when strategizing long-term research directions for Generative AI. For instance, the \emph{continually evolving} nature of cybersecurity knowledge bases means that GenAI models must be continually adapted~\citep{luo2023empirical, gururangan2020dontstop, hu2021lora, shi2024continualsurvey}. In cybersecurity, adaptation isn't just an advantage---it's essential for survival.

Beneath this complexity, there lies an equally important and defining goal for practitioners: precision. As the digital landscape grows in scale and complexity, the pressure on defenders is not simply to know more, but to act with sharp focus. Precision is what separates actionable security from overwhelming background noise, and what allows technical solutions to translate into measurable risk reduction. In other words, the essence of precision is not only in gathering threat intelligence or vulnerability data, but in carefully discerning which items are \textit{critical given context}, and what exact, feasible steps are required to mitigate them.
    
In the pre-GenAI era not so long ago, it was virtually impossible for an individual or an organization to make sense of all the constantly emerging cybersecurity data emanating from countless sources around the world. Although we had the advantage of several communities, forums, and organizations actively contributing towards gathering knowledge and advancing the science to intelligently tackle sophisticated cyberthreats, the siloed nature of these resources limited our ability to achieve \emph{precision in cybersecurity}~\citep{huang2024ctikg, fieblinger2024actionable}.
        
\begin{figure}[t]
  \centering
\includegraphics[width=0.8\linewidth]{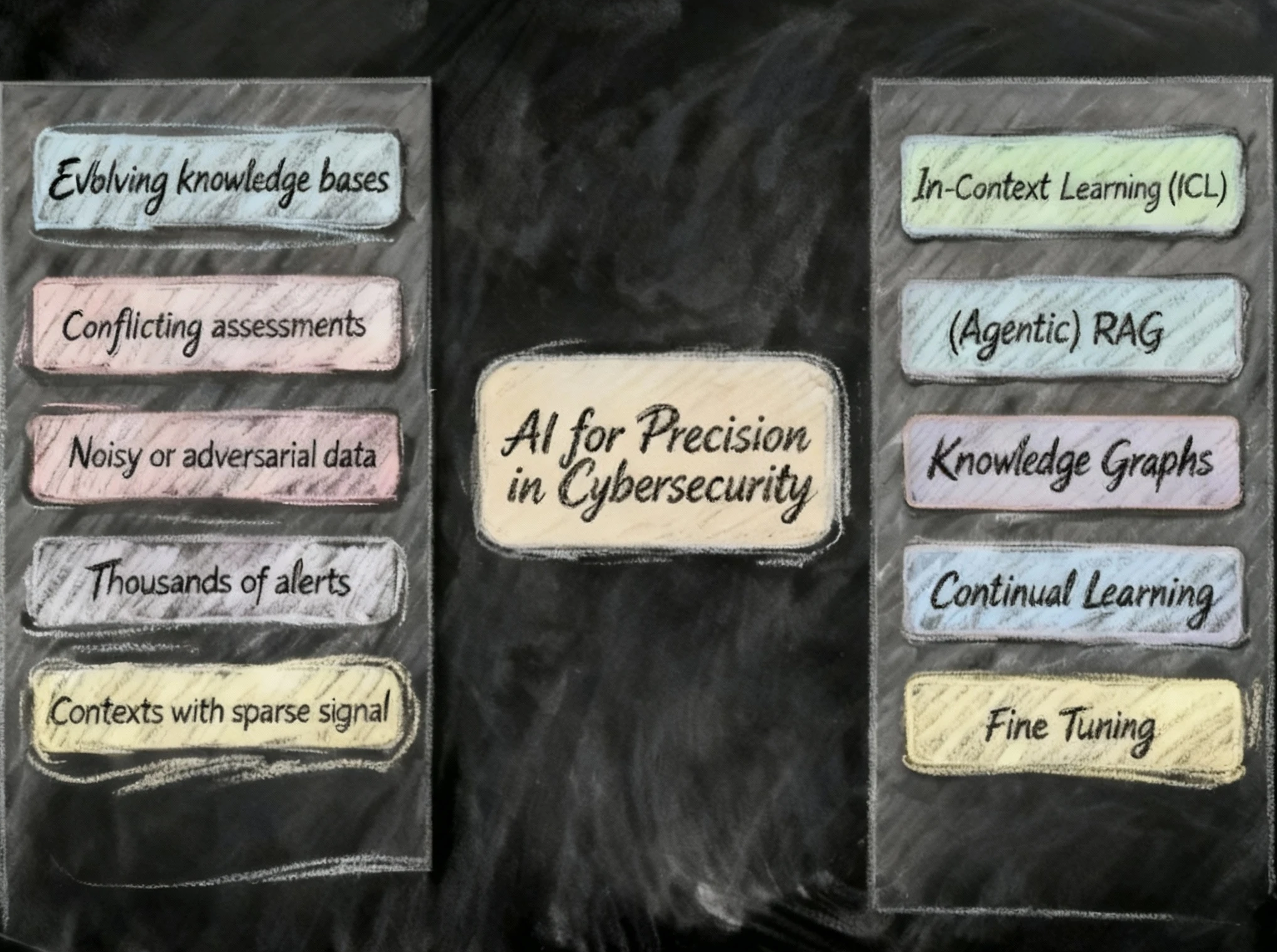}
  \caption{AI for Precision in Cybersecurity: An overview of the challenges and adaptation mechanisms discussed in this paper. On the left, we show core obstacles—such as evolving knowledge bases, conflicting and noisy assessments, countless alerts, and sparse-signal contexts—that make precision in cybersecurity both essential and difficult. On the right are the main approaches for continual adaptation: inference-time adaptation (with in-context learning, retrieval-augmented generation, and knowledge graphs), and model-level adaptation (through continual learning and fine-tuning). Together, these orthogonal levers enable GenAI systems to transform massive, fragmented cybersecurity data into precise, context-aware actions in this never-ending, rapidly changing domain.}
  \label{fig:intro-precision}
\end{figure}
    
Concretely, production environments routinely confront thousands of identified vulnerabilities, yet only a handful are immediately exploitable or relevant for a specific enterprise's operations. The ability to efficiently sift through this sea of alerts, surface the issues that actually warrant attention, and recommend specific remediations that are both effective and non-disruptive, is at the core of achieving precision.

For instance, if remediation involves package upgrades or downgrades, ``precision" means recommending a version that is not only secure, but also stable and compatible with all other dependencies~\citep{moderne2023dependency, wawand2025dependency}---avoiding a cascade of breakages in production and enabling fast, safe deployment~\citep{legit2025dependency}. The challenge is compounded by the reality that a majority of security fixes require minor or major version updates rather than simple patch bumps~\citep{moderne2023dependency}, and that transitive dependencies can introduce hidden vulnerabilities deep in the dependency tree. Thus, precision transforms overwhelming and ambiguous alerts into clear, prioritized guidance for confident, impactful action.
    
It is only with the advent of large language models (LLMs) that the limits of traditional pipelines—unable to integrate, reason over, and prioritize across siloed or noisy cybersecurity data—become fully apparent. LLMs offer the ability to “find the needle in the haystack”~\citep{kamradt2023needle, liu2023lostinthemiddle, hsieh2024calibrating_position}, extracting precise answers from fragmented and pressured knowledge at scale. However, even this breakthrough comes with pitfalls: modern LLMs struggle with reasoning as input length grows, and their performance may degrade with overly long or noisy prompts—even when all necessary information is in the context window~\citep{shi2025context_length_hurts, chroma2025context_rot}.
    
Another important advantage of an LLM is its innate ability to implicitly reason about different pieces of information (in the form of text, table, etc.) from disparate sources that it has seen during pretraining, together with any additional information provided in the prompt (for example, a diverse set of CVSS analyses of a vulnerability by NVD, Red Hat, Ubuntu, Amazon Linux, Bitnami, GHSA, etc.)~\citep{wei2022chainofthought,kojima2022large}. In an agentic framework, the capabilities of LLMs as AI agents are further enhanced to perform complex tasks as a team, often outperforming single-model approaches through role-playing and collaboration~\citep{li2023camel, hong2023metagpt, qian2023chatdev, tran2025multiagent_survey}.

Notably, self-correction mechanisms (both intrinsic and via post-hoc agent orchestration) offer promise for refining LLM outputs, though their performance varies and the danger of ``compounded error propagation" in agentic, multi-step cyber reasoning remains~\citep{li2025correctbench,madaan2023selfrefine,huang2024llm_selfcorrect_limits}.
Despite the smart engineering of AI agents through a variety of agentic frameworks~\citep{langchain2024, wu2023autogen}, the cybersecurity domain presents unique challenges~\citep{xu2025agentic, deng2024pentestgpt, shahriar2025agentic_security}.
    
Considering the constantly accumulating and ever-evolving knowledge bases in cybersecurity (for example, known exploits of existing vulnerabilities, CVSS or EPSS analysis~\citep{first2023epss}, security measures by enterprise operating systems or clouds), a pre-trained LLM despite its super intelligence and vast knowledge is not sufficient on its own. Thus, for precision in the ever-changing world of cybersecurity, practitioners must \emph{continually adapt AI}. As we discuss next, there are various ways to continually adapt AI. One can explicitly perform continual adaptation of the AI models themselves via deep continual learning~\citep{meng2022locating, meng2022mass, ke2023continualpretrain, kirkpatrick2017ewc}. Alternatively, AI models can consume the continuously evolving information in knowledge bases through well-known techniques such as Retrieval Augmented Generation (RAG)~\citep{lewis2020rag, borgeaud2022retro, jiang2023active,yang2024gptvsretro,wang2023shall,gao2024retrieval,shao2024scaling}. In the latter case, RAG pipelines must be aware of the dynamic nature of the knowledge bases and should be optimized accordingly~\citep{guu2020realm, asai2024selfrag}. In the following, we discuss these techniques not only from the perspective of continually adapting AI but also in general for their relevance to AI pipelines for cybersecurity. We discuss the pros and cons of these techniques along with strategies to make the techniques more robust, especially against inherent noise in the datasets~\citep{yoran2024robustrag, karpukhin2020dense}.

Bridging research and practice, this paper illuminates how practitioners must proactively harness both retrieval- and model-level approaches—and structure knowledge bases with auditable provenance and interrelations—if GenAI systems are to match the speed and sophistication of today’s threat landscape.

Broadly speaking, there are two levers we can pull: we can change \emph{what} the model sees at inference time (through better retrieval and context construction), or we can change \emph{the model itself} (through continual learning). In Section~\ref{sec:icl-rag}, we focus on the former, examining in-context learning, RAG, and knowledge graphs. In Section~\ref{sec:continual}, we turn to the latter, discussing when and how to continually pretrain LLMs for cybersecurity.

\begin{keypointbox}[Key ideas]
\begin{itemize}
    \item Cybersecurity is a \textbf{never-ending, fast-changing game}: problems, tools, and threat landscapes can become obsolete in months.
    \item \textbf{Precision} is central: defenders must map massive, noisy data (CVEs, exploits, advisories, telemetry) to a small set of \emph{context-aware, feasible actions}.
    \item Pre-GenAI pipelines could not fully exploit fragmented security knowledge; LLMs can now \textbf{integrate and reason} over heterogeneous sources.
    \item However, LLMs alone are not enough: \textbf{knowledge bases change faster} than static models; supply-chain and ecosystem risks evolve continuously.
    \item The rest of the paper frames GenAI for cybersecurity around two orthogonal levers:
    \begin{enumerate}
        \item \textbf{Inference-time adaptation}: ICL, RAG, and knowledge graphs.
        \item \textbf{Model-level adaptation}: continual (unsupervised) pretraining / fine-tuning.
    \end{enumerate}
    \item In short: Precision and resilience demand not just smarter AI, but smarter adaptation pipelines that can keep up with rapid change.
\end{itemize}
\end{keypointbox}

\section{In-Context Learning and Retrieval Augmented Generation}
\label{sec:icl-rag}

To address the challenge of accommodating new and rapidly changing knowledge in generative AI pipelines, the most common, simple, and practical approaches are In-Context Learning (ICL) and Retrieval Augmented Generation (RAG)~\citep{brown2020fewshot, lewis2020rag}. Both techniques leave the model parameters fixed, but modify the prompt so that the model can condition on relevant, often very recent, information. For the cybersecurity domain in particular, ICL and RAG can be further strengthened in terms of efficiency and robustness with the use of knowledge graphs~\citep{pan2024graphrag, edge2024graphrag_microsoft}.
%
Designing robust real-world pipelines most often means hybridizing ICL, RAG, and knowledge-graph-enhanced retrieval, orchestrating together dense retrieval, structured filtering, and LLM-centric decision agents for evaluable and resilient results.
    
\begin{figure}[t]
  \centering
\includegraphics[width=0.8\linewidth]{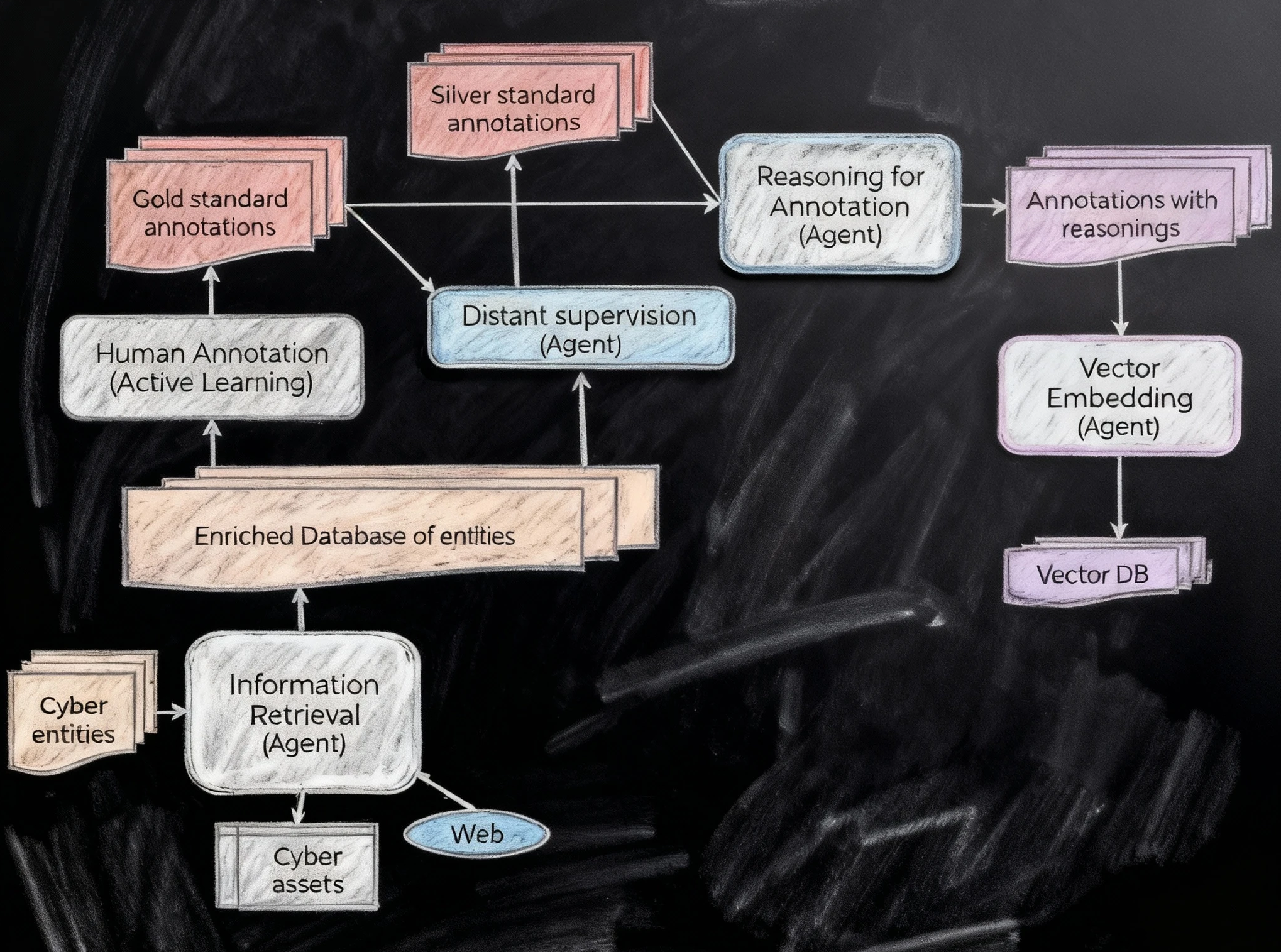}
\caption{
  Illustration of a typical data pipeline for cybersecurity sketched for this paper’s context. The process begins at the lower left, where an information retrieval agent gathers cyber entities and assets from internal sources and the web, consolidating them into an enriched database of entities. High-quality, manually-curated “gold standard” annotations are provided by human experts, while automated agents generate broader “silver standard” annotations using distant supervision techniques. Reasoning agents further augment these annotations with explanatory rationales to enhance interpretability. The resulting set of annotated and explained entities is transformed into machine-usable vector representations by a vector embedding agent and stored in a vector database. This integrated approach balances expert knowledge with scalable automation and reasoning, supporting the pipeline’s ability to adapt and maintain precision in the complex, evolving cybersecurity landscape.
}
  \label{fig:data_pipeline}
\end{figure}

In-context learning refers to providing labeled examples as training data as part of the prompt itself~\citep{brown2020fewshot, min2022rethinking_icl}. For certain tasks, a few labeled examples can suffice as the training set for an LLM to improve its accuracy, popularly referred to as Few-Shot Learning~\citep{wei2022flan}. When we retrieve only those labeled examples that are relevant to a specific query and include them in the corresponding prompt, we effectively combine RAG with ICL (often called RAG-ICL). Of course, RAG is a more general concept and is not limited to labeled examples. The general idea is to obtain all the information relevant to a query as an additional context to be processed by an LLM as part of the prompt. The retrieval is performed using semantic embeddings of text~\citep{reimers2019sbert, karpukhin2020dense}, or more recently using AI agents, possibly leveraging knowledge graphs for efficiency and robustness to noise. Domain-specific embedding models, such as those fine-tuned for cybersecurity~\citep{cybert2021ranade, securebert2025}, have demonstrated superior performance over general-purpose models in cybersecurity-specific tasks including threat classification, vulnerability assessment, and semantic search. 

\begin{keypointbox}[Mental model: ICL, RAG, and knowledge graphs]
\begin{itemize}
    \item \textbf{ICL}: Feed the model a few labeled examples inside the prompt to allow it to infer the pattern on the fly.
    \item \textbf{RAG}: Retrieve relevant text from external corpora (e.g., CVEs, advisories, internal logs) and attach it to the prompt.
    \item \textbf{RAG + ICL}: Retrieve \emph{labeled} examples (e.g., past triage decisions) to guide the model on similar new cases.
    \item \textbf{Knowledge graphs}: Add structure (nodes/edges for CVEs, packages, TTPs, etc.) so retrieval is constrained by domain relationships, not only by embedding similarity.
    \item All these are \emph{inference-time} levers: the model weights stay frozen; only the prompt/context changes.
\end{itemize}
\end{keypointbox}

\subsection{Noisy Context for RAG}

What if the information retrieved as the context for a query is not relevant but rather irrelevant noise—text snippets that may degrade the accuracy of the answer instead of improving it? Such scenarios are more common in practice than one would imagine. This is the fundamental limitation of RAG, in addition to the computational overhead of lengthy prompts and the laborious engineering efforts involved in maintaining RAG pipelines~\citep{gao2024rag_survey}. In real-world settings, context noise can have an outsized impact on LLM reliability, especially for high-risk decisions. Research has demonstrated that retrieval augmentation can paradoxically hurt performance when irrelevant context is introduced~\citep{yoran2024robustrag}. Specifically, as the number of retrieved passages increases, LLM accuracy often plateaus or even declines due to attention disruption from irrelevant content~\citep{shi2025layer_knowledge}. 

Finding the pieces of text relevant to a query remains a challenging problem, often presenting a trade-off between precision (how many of the retrieved text snippets are actually relevant to the query) and recall (are all the text pieces of information required for answering the query retrieved?), especially for challenging domains like cybersecurity or biology~\citep{karpukhin2020dense}. In RAG pipelines, high recall from the retriever is typically prioritized over precision because missing critical context can prevent the generator from formulating correct responses, even if the few retrieved documents are highly precise. The trade-off manifests practically: retrieving more documents improves recall linearly but introduces noise that disrupts attention~\citep{shi2025layer_knowledge}. Unfortunately, for a domain like cybersecurity, high recall at the cost of low precision is not acceptable. For instances where precision can't be ensured, it is suggested to bring a human expert in the loop.

This trade-off between precision and recall is further pronounced for scenarios where a query is highly technical, requiring searching of a small number of relevant texts through a very large corpus. For such scenarios, semantic embedding-based retrieval of the relevant text is certainly noisy for various technical reasons~\citep{reimers2019sbert}. Here, it is worth noting that semantic embeddings are an older family of techniques with fundamental limitations, which helped motivate the development of LLMs as a breakthrough for generative AI.

\textbf{Example:}
Suppose you want to retrieve information about recent SSRF (Server-Side Request Forgery) vulnerabilities specifically in AWS Lambda. If you rely solely on text-based embedding similarity, you may inadvertently retrieve results for unrelated AWS services or generic blog posts about SSRF affecting different cloud platforms. Such irrelevant matches occur because the retrieval mechanism lacks precise understanding of service boundaries or context, underscoring the need for domain-specific filters or structured context.

This is why a more efficient approach is to leverage an LLM for the problem of retrieval itself, namely \emph{Agentic RAG}, which is becoming more practical with new LLMs able to process prompts with millions of tokens~\citep{anthropic2024claude, gemini2024million}.

\begin{keypointbox}[Key pitfalls of RAG in cybersecurity]
\begin{itemize}
    \item \textbf{More context is not always better}: beyond a point, extra retrieved passages introduce noise and can \emph{reduce} accuracy.
    \item \textbf{Precision vs.\ recall}: security workloads often need \emph{high precision}, but most RAG stacks are tuned for \emph{high recall}.
    \item Highly technical queries (e.g., specific cloud misconfigurations) are especially vulnerable to noisy retrieval from large corpora.
    \item Static embedding-based retrieval alone is brittle; it struggles when the signal (true relevant text) is sparse and buried in a huge corpus.
    \item When precision cannot be guaranteed, \textbf{human-in-the-loop review} is essential—especially for decisions that affect production systems.
    \item Build in escalation paths from RAG pipelines to expert review, and monitor for drift in retriever behavior.
\end{itemize}
\end{keypointbox}

\subsubsection{Agentic RAG}
One may ask why we should need RAG in the first place if an LLM can already process prompts with millions of tokens~\citep{liu2023lostinthemiddle}. In practice, the LLMs used as agents to retrieve the relevant texts can be different from the primary one employed for answering the query. The reasoning abilities of an LLM can vary significantly when processing millions of tokens as input vs. a few thousand tokens~\citep{hsieh2024calibrating_position}. This phenomenon, sometimes termed ``Context Degradation Syndrome," manifests as gradual breakdown in coherence during long-running conversations~\citep{howard2024cds}, with performance degradation occurring even when models can perfectly retrieve relevant evidence~\citep{shi2025context_length_hurts, chroma2025context_rot}. So, it would still be wise to have a RAG-specialized AI agent process a corpus as an input to select text pieces relevant for a query (finding a needle in the haystack) and then have another AI agent process the selected pieces of text as the relevant context to answer the original query. 

To reduce the cost of having to process millions of tokens by the retrieval agent, one can use semantic embeddings as a preprocessing step to remove those text snippets which are surely irrelevant to the query. Besides the cost of potentially processing millions of tokens for a single query, Agentic RAG has some practical limitations, especially when a large corpus is input as a prompt such as:
\begin{itemize}
    \item Potentially ignoring a large fraction of the prompt (many pieces of text in an input corpus which could be relevant to the query), thus possibly missing the sought needle in the haystack~\citep{liu2023lostinthemiddle};
    \item Hallucination of data is a prominent problem which is exacerbated as an input corpus grows in size or diversity of its content~\citep{huang2023hallucination_survey, ji2023hallucination_survey};
    \item The model may produce over-generalized output that misses niche details relevant to the query and ignores the richness of the domain, particularly when needle-question similarity decreases~\citep{chroma2025context_rot}.
\end{itemize}

\textbf{Illustrative failure:}
Imagine you ask a RAG pipeline, ``Which Red Hat updates fixed the log4shell vulnerability in JBoss?" If context selection is imprecise, it might return unrelated bug reports about general log messages, or overlook fixes because they were published under a different advisory headline. As a result, defenders could miss critical patches or review irrelevant information—showing the risk of relying solely on surface-level retrieval.

\subsubsection*{}

Unfortunately, there is no clear winner when it comes to choosing one of the known techniques in the literature for the real-world efficiency of RAG~\citep{gao2024rag_survey}. In practice, depending upon the task at hand, it is a (sparse) mixture of different techniques at play. On the other hand, one thing that rarely goes wrong even in the era of AI is the use of domain knowledge, which often brings higher efficacy, robustness, and interpretability in AI pipelines. Along these lines, we discuss one such practical case of leveraging domain knowledge for RAG, namely, the use of knowledge graphs to structure and constrain retrieval.

\begin{keypointbox}[Agentic RAG in practice]
\begin{itemize}
    \item Use a \textbf{retrieval agent} (possibly a smaller LLM) to:
    \begin{itemize}
        \item read large, messy corpora,
        \item shortlist candidate passages,
        \item and construct a focused context.
    \end{itemize}
    \item Use a \textbf{separate answer agent} (possibly a larger LLM) to reason deeply over the curated context.
    \item Combine embeddings + LLM agents:
    \begin{itemize}
        \item Embeddings to prune \emph{obviously} irrelevant material cheaply.
        \item Agents to make \emph{finer-grained} decisions about what is truly relevant.
    \end{itemize}
    \item Be aware of failure modes:
    \begin{itemize}
        \item Ignored parts of long prompts.
        \item Over-generalized answers that miss niche but critical details.
        \item Hallucinations growing with corpus size/diversity.
    \end{itemize}
\end{itemize}
\end{keypointbox}

\subsection{Leveraging Knowledge Graphs for RAG}
For the cybersecurity domain, in many scenarios, one can avoid being completely reliant upon semantic embeddings or AI agents to retrieve contextual information relevant for a query~\citep{pan2024graphrag}. This is because cybersecurity knowledge is naturally expressed in an implicit or explicit knowledge graph with nodes representing entities such as vulnerabilities, software artifacts, etc.~\citep{mitre2024attack}. 

Recent work has demonstrated the value of unifying CVE, CWE, and CPE databases into cohesive knowledge graphs~\citep{shi2023uncovering_relations}.
Knowledge graphs act as a precision filter—enabling targeted discovery of relationships (e.g., exploits-to-assets, vulnerabilities-to-packages, TTPs-to-mitigation) and facilitating interpretable, auditable explanations for AI-driven analyses.

As an example, suppose that a query concerns the log4j package. Let us assume that there is a node in the knowledge graph that represents the log4j package. This means that all pieces of information related to the log4j package should be attributes of the same node or should be retrieved from its neighboring nodes. Here, as typical for real-world knowledge graphs, one can safely assume that there are multiple types of nodes in the same graph associated with different types of entity (including a node of nonentity type). The links between different nodes are established based on the semantics of the node attributes, as well as the observed connections between different entities. 
    
For instance, a node associated with the log4j package should be associated with all the nodes representing the software vulnerabilities contained by log4j such as CVE-2021-44228, CVE-2021-45046, CVE-2021-45105, CVE-2021-44832, CVE-2021-4104, etc. Similarly, the node for log4j can be linked to other package nodes in the knowledge graph that are similar in their characteristics (say, share common vulnerabilities). To take it one step further, categorical features of a vulnerability like a CVSS vector (attack vector, attack complexity, privileges required, etc.) or its type (RCE, DoS, SSRF, etc.) can be leveraged to group similar vulnerabilities, thus connecting them to each other in the knowledge graph~\citep{first2024cvss4}. 

\textbf{Additional Example:}
Suppose an analyst wants to identify all known supply chain attacks targeting JVM-based logging libraries—such as both \texttt{log4j} and \texttt{logback}. Using only embedding-based textual similarity, relevant vulnerabilities may be missed because these libraries might not explicitly mention each other or share overlapping descriptive text. However, a knowledge graph that includes explicit type and dependency relationships (e.g., nodes tagged as JVM logging library or linked via \emph{depends-on} or \emph{category} edges) allows the analyst to easily traverse from the broader JVM-based logging library node to all constituent libraries (including log4j and logback) and then retrieve associated vulnerabilities, even if their textual descriptions differ. This structured approach provides comprehensive and precise results that embedding similarity alone would likely overlook.

Such graph-structured data also support explainability—an LLM can trace links between advisory nodes, package nodes, and exploit nodes, providing holistic justifications for risk scores or mitigation recommendations.
    
Thus, in consideration of the possible inefficiency of semantic embeddings or AI agents in retrieving the relevant content for a given query, named entities such as package names, vulnerability IDs, or even the general structural nature of cybersecurity knowledge come in handy wherever possible~\citep{edge2024graphrag_microsoft}. It is worth noting that knowledge as a whole or even in pieces in the cybersecurity domain can be represented as a knowledge graph that can be cleverly used to build robust and cost-effective RAG pipelines. To clarify further, in most cases, a knowledge graph does not need to be built explicitly for RAG.

Given a knowledge graph as one of the primary tools for RAG, semantic embeddings can also be utilized wherever applicable. For example, in addition to the observed links between nodes in a knowledge graph, semantic embeddings can be computed for textual attributes of nodes to reveal the hidden semantic links between nodes, which is a classical link prediction problem in the literature on social networks and graph representation learning~\citep{wang2024kge_survey, tgrail2023inductive_kge}. 

\begin{keypointbox}[Using knowledge graphs for RAG]
\begin{itemize}
    \item Cybersecurity data naturally fits a graph:
    \begin{itemize}
        \item Nodes: CVEs, CWEs, CPEs, packages, OS images, ATT\&CK techniques.
        \item Edges: \emph{affects}, \emph{depends-on}, \emph{shares-vulnerability-with}, \emph{similar-CVSS-profile}, etc.
    \end{itemize}
    \item For a given query (e.g., log4j), start from relevant nodes and their neighborhoods instead of scanning the full corpus.
    \item Use graph structure to \textbf{constrain retrieval}:
    \begin{itemize}
        \item Only fetch vulnerabilities attached to affected products.
        \item Only fetch exploit notes connected to the CVE nodes in question.
    \end{itemize}
    \item Embed the graph (or node attributes) if you need similarity search, but use the graph to \textbf{filter and prioritize} candidates.
\end{itemize}
\end{keypointbox}

\subsection{Tune Semantic Embeddings using Knowledge Graphs}
Moreover, a strategic approach is to tune the semantic embeddings using the knowledge graph~\citep{wang2024kge_survey}. In other words, knowledge graphs and semantic embeddings can have a symbiotic relationship. As an outcome from such tuning of semantic embeddings and inference of hidden links in a knowledge graph, depending upon preference or task at hand, one can utilize either of the two or both for a RAG pipeline. Appropriately tuned semantic embeddings of entities are supposed to encapsulate all the domain knowledge of the graph. They can be useful resources along with the knowledge graph, for the cybersecurity community at large. 

This synergy is bidirectional: Knowledge graphs provide the backbone for constraining and filtering embedding-based retrieval, whereas learned embeddings can accelerate link prediction, node clustering, and support generalization to unseen or fuzzily described entities in cyber queries.
    
From an AI research perspective, our long-term strategy should be to publicize such tuned semantic embeddings of entities (vulnerabilities, packages, etc.), just as there are public releases on exploitability or impact of vulnerabilities.

Tuning of semantic embeddings can be accomplished by either tuning embedding-only models (for example, BERT~\citep{devlin2019bert}) or via tuning of the full-fledged generative models (LLMs)~\citep{reimers2019sbert}. In the literature of deep learning, in particular graph neural net learning on top of knowledge graphs, it is a well known problem to learn node embeddings (and edge embeddings) as representations of node attributes as well as of their links/edges/relations to other nodes in the knowledge graph~\citep{kipf2017gcn, hamilton2017graphsage}. Often, the LLM and the graph neural network are fine-tuned jointly~\citep{tgrail2023inductive_kge}.

\textbf{Practical suggestion:} Publish open-source, domain-tuned embeddings (for example, embedding vectors for every known CPE, CVE, or MITRE ATT\&CK technique~\citep{mitre2024attack}) to accelerate broader innovation in the community. Public benchmarks on downstream cybersecurity tasks would catalyze progress in AI-for-security pipelines, similar to benchmarks in NLP and CV.

The past couple of decades in cybersecurity witnessed contributions towards structuring the knowledge of cybersecurity (for example, MITRE frameworks~\citep{mitre2024attack}, EPSS~\citep{first2023epss}, CVSS~\citep{first2024cvss4}, etc.). Hopefully, in this decade, we will contribute open-source embeddings of entities to enable highly accurate yet low-cost AI pipelines for our collective goal of precision in cybersecurity. 
        
\begin{keypointbox}[Embeddings--graph synergy: a recap]
\begin{itemize}
    \item \textbf{Knowledge graphs} give you:
    \begin{itemize}
        \item Explicit structure and relationships.
        \item Transparent, auditable reasoning paths (e.g., ``this CVE affects these packages'').
    \end{itemize}
    \item \textbf{Embeddings} give you:
    \begin{itemize}
        \item Fuzzy similarity (``these advisories \emph{feel} alike'').
        \item Generalization to new text and unseen combinations of terms.
    \end{itemize}
    \item \textbf{Tune embeddings using the graph}:
    \begin{itemize}
        \item Encourage nodes that are close in the graph to be close in embedding space.
        \item Improve link prediction and RAG retrieval quality simultaneously.
    \end{itemize}
    \item Long-term vision: \textbf{open, domain-tuned embeddings} for entities like CVEs, CPEs, and ATT\&CK techniques, shared across the community.
\end{itemize}
\end{keypointbox}

The techniques discussed so far focus on improving \emph{retrieval} and \emph{context construction} around a largely static base model. In many cases this suffices, but in others it is more natural and efficient to update the model itself. We now turn to this second adaptation lever: continual lifelong learning of LLMs.

\begin{figure}[t]
  \centering
\includegraphics[width=0.8\linewidth]{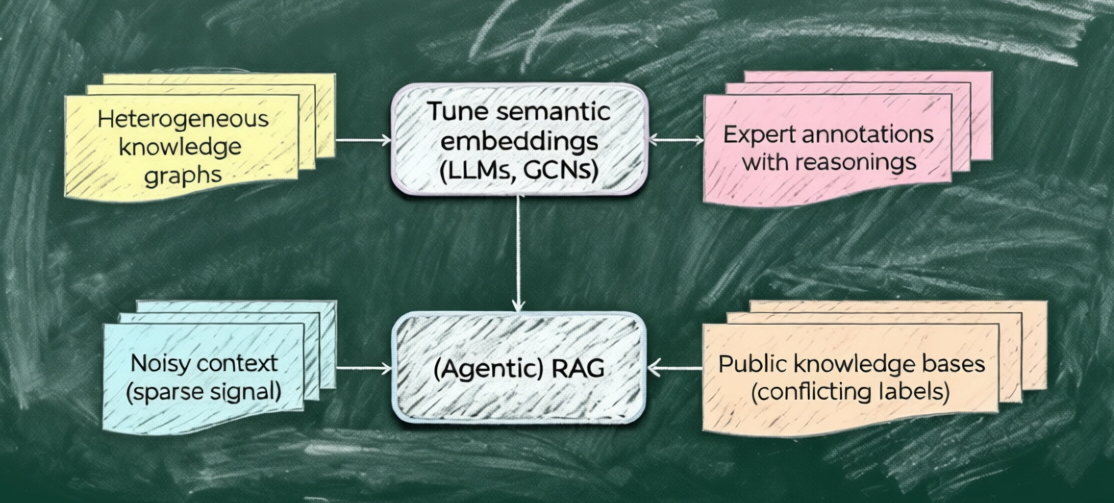}
  \caption{
  %
  This figure illustrates how a GenAI pipeline in cybersecurity can bring together structured knowledge graphs—which map relationships among entities like CVEs, packages, and attack techniques—and expert-annotated examples to tune semantic embedding models (LLMs, potentially with GCNs). These embedding models, enriched with both human insight and graph structure, are then used alongside real-world context and public knowledge bases to support robust Retrieval-Augmented Generation (RAG). Agentic RAG, leveraging LLM agents, can further improve precision and context selection in the face of noisy, conflicting, or fast-changing security information.
}
  \label{fig:rag}
\end{figure}

\section{Continual Lifelong (Deep) Learning of LLMs}
\label{sec:continual}
Although the above discussed strategies for RAG can be sufficient for a majority of the RAG pipelines, there are scenarios where continual (deep) learning of LLMs (aka continual pretraining) is too advantageous to be ignored~\citep{ibrahim2024continual_pretrain, yildiz2024continual_insights, ke2023continualpretrain}. Before we discuss such scenarios, there are a few key points worth noting about the continual learning of LLMs as proposed here.

        
\subsection{Key Characteristics of Continual Learning}

Continual learning is a broad broad technique which can be approaches in several possible ways. In the following, we discuss key traits of continual learning we propose for cybersecurity domain based on our own experience of continual learning of LLMs for various tasks. Depending upon tasks at hand, some of these traits may or may not be applicable for a practitioner.
    
\subsubsection{Unsupervised objective for learning}
    
We solely advocate for unsupervised (continual) learning of LLMs, rather than supervised learning or reinforcement learning~\citep{ouyang2022instructgpt}. Unsupervised learning has a lower computational cost and is often more robust than its counterparts. 
    
Contrary to prior beliefs, unsupervised learning can be powerful enough to enable LLMs to memorize as well as reason about the new knowledge from training corpora, as is evident from pretrained LLMs which not only remember the knowledge from training corpora, thus serving as a new generation of knowledge bases (and not just as chat agents that hallucinate facts), but also demonstrate (super intelligence) advanced reasoning skills in their responses (reflecting intelligent assimilation of knowledge coming from disparate sources of text in training corpora)~\citep{brown2020fewshot, touvron2023llama2}. 
    
Naturally, what has been accomplished during LLM pretraining can be replicated on a smaller scale by continuing LLM pretraining on corpora from the cybersecurity domain~\citep{gururangan2020dontstop}. There are some practical tricks to make such learning efficient in practice, often varying as per a domain and tasks at hand; as an example, hyperparameters for the learning including regularization of the cross-entropy objective through masking are tuned for the cybersecurity domain. 
    

\textbf{Example:}
For instance, by continually fine-tuning a cybersecurity LLM on daily feeds of new vulnerabilities and attacker techniques, the model can begin to predict which types of exploits may become widespread or recommend proactive mitigations—even if these threats have not yet been widely reported in security news. This adaptation is achieved without explicit human labeling of each incident.

\subsubsection{Unsupervised fine tuning via LoRA}
Perhaps surprisingly to some of us, continual pretraining of LLMs can be performed with Low-Rank Adaptation (LoRA) as well~\citep{hu2021lora}, i.e. unsupervised fine-tuning of LLMs (usually, fine-tuning is by default assumed to be supervised learning, which is not always the case). 
    
Thus continual pretraining-based fine tuning of LLMs can be performed with a small set of GPUs in contrast to tens of thousands of GPUs utilized for pretraining LLMs, especially if it is task-driven utilizing medium sized training corpora (in the order of a few hundred million tokens)~\citep{dettmers2023qlora}. The preference for fine-tuning-based unsupervised learning (continual pre-training) is not solely driven by the compute cost but to mitigate the possibility of forgetting the existing knowledge and intelligence of a pre-trained LLM~\citep{kirkpatrick2017ewc, mccloskey1989catastrophic}. 
    
Despite the great success of deep learning for generative AI, continual learning of deep neural nets remains one of the fundamental problems, as it presents the trade-off between forgetting old knowledge (tasks) for attaining new knowledge (tasks)~\citep{kirkpatrick2017ewc}. Large language models do have the upper hand on this trade-off for their capacity (large number of parameters) to keep on learning new knowledge without forgetting the old ones. However, as we understand, LLMs are already pre-trained to the saturation point (for example, Llama 3.3 70B~\citep{touvron2024llama3}). In such cases, the trade-off becomes more severe, requiring advanced techniques for efficient (full fledged) continual learning~\citep{ibrahim2024continual_pretrain}. On the other hand, fine-tuning relaxes such trade-offs as recent studies suggest~\citep{hu2021lora, dettmers2023qlora}, thus being a safer choice from both a theoretical and a practical standpoint.

\textbf{Tip:} Fine-tune your primary LLM (via LoRA or QLoRA~\citep{dettmers2023qlora}) on domain-specific corpora monthly or quarterly to keep up with adversarial TTP evolution; always validate on a public benchmark to avoid catastrophic forgetting.
Monitor impact on downstream tasks, using stable benchmark suites to track for regression or drift.

\subsubsection{Not a Jack of all trades}
Furthermore, from our experience, we do not recommend (unsupervised) fine-tuning of a single all-purpose LLM for the cybersecurity domain. Rather, in our pipelines, we have a basket of fine-tuned LLMs (with unsupervised learning) for a diverse set of tasks. 
    
While in theory it is possible to have a single LLM fine-tuned on corpora from all the tasks, it would make the process of learning more challenging for similar theoretical reasons as cited above in favor of fine tuning~\citep{shi2024continualsurvey}. (Unsupervised learning-based) Fine-tuning for an individual task is a more flexible and robust approach. 
    
Having said that, there is no golden rule and there are real life scenarios that demand having a single jack-of-all-trades LLM which can be hosted 24x7 for responding to any kind of query (prompt) across different tasks. For such cases, we recommend fine-tuning an LLM more carefully, being aware of the literature and the subtle tricks, rather than treating fine-tuning libraries (such as torchtune) as black boxes.
    
When fine tuning models specific to tasks, one of the relevant strategies is to merge models, where task-specific expert models (fine-tuned via LoRA) can be combined without retraining~\citep{nvidia2024model_merging}, or organized in Mixture-of-Experts (MoE) architectures where specialized experts are dynamically engaged based on task requirements~\citep{tensorops2025moe, icml2024moe_survey}.
    %
    
\textbf{Rule of thumb:} For critical, automated workflows (such as vulnerability prioritization or automated mitigations), specialty fine-tuned models offer greater control and auditability over monolithic jack-of-all-trades LLMs. Use generalists for broader exploratory or semi-structured interactive use cases.

\subsubsection{Why unsupervised objective suffice for learning from labeled datasets?}
For most of the back-end tasks in cybersecurity, it is the knowledge and intelligence (reasoning skills) of an LLM that is of primary relevance~\citep{wei2022chainofthought}. Let us consider the task of classifying the role of a software package from categories such as server, client, library, driver, etc. Suppose that we have annotations on the labels provided by security researchers for those packages for which the classification is ambiguous and challenging. When fine-tuning an LLM with the label annotations, our objective is not for the LLM to memorize the labels of the packages. Rather, the primary goal is that an LLM should understand the reasoning (expressed in language unlike labels) behind those chosen labels. It is the reasoning that will translate (generalize) to classification of unseen (test) software packages, not to memorizing the actual labels. (In fact, in most cases, reasoning for the labels can be reliably obtained by an LLM itself~\citep{kojima2022large}.) For an LLM to attain and encapsulate this knowledge in its weight parameters, unsupervised fine-tuning suffices.
    
Remember, reasoning is expressed in the natural language itself, and unsupervised learning is sufficient to understand and learn any form of language and its underlying deep reasoning as evident from the super intelligence of LLMs. Techniques like supervised fine tuning or reinforcement learning are meant for more subtle goals such as instruction following~\citep{wei2022flan, ouyang2022instructgpt, zhou2023lima}, alignment to human preferences~\citep{ouyang2022instructgpt}, etc. These goals are relatively rare in cybersecurity and, in our experience, the existing skills of pre-trained LLMs are sufficient in this matter unless those skills were forgotten in the process of continual learning. 

\textbf{Example:} Feed the LLM sentences explaining why a package is a ``server" (e.g., ``It listens for inbound HTTP traffic on a well-known port."), rather than just the label. The LLM, even unsupervised, will learn such features more robustly and extrapolate to unseen packages.
    
\begin{keypointbox}[Design principles for continual learning in cybersecurity]
\begin{itemize}
    \item Prefer \textbf{unsupervised continual pretraining} on fresh domain text (advisories, TTPs, exploit write-ups) over costly supervised/RL pipelines.
    \item Use \textbf{LoRA/QLoRA} to:
    \begin{itemize}
        \item keep compute costs manageable,
        \item reduce catastrophic forgetting risks on the base model.
    \end{itemize}
    \item Maintain a \textbf{basket of specialist models} (per task or per cluster of tasks) rather than one giant jack-of-all-trades model, unless you have strong reasons and careful evaluation.
    \item Let labeled datasets influence training primarily through \textbf{rationales and explanations}, not just raw labels: we want to transfer reasoning patterns.
    \item Reserve instruction-tuning/RLHF mainly for interaction style, preference alignment, and reasoning; rely on unsupervised learning to ingest factual and structural knowledge.
\end{itemize}
\end{keypointbox}

\subsection{When to Continually Learn LLMs?}
Although in theory some LLMs can process a prompt of millions of tokens~\citep{gemini2024million}, their efficacy for real-world scenarios does suffer as the prompt size increases even beyond a few thousand tokens, especially when precision and objectivity are of importance for the outcome of an AI pipeline, and when latency and cost constraints limit how much context can be fed per query~\citep{liu2023lostinthemiddle}. In the following, we discuss various scenarios when it is worth considering continual pretraining of LLMs. 
    
\subsubsection{When RAG may fail with large and noisy corpora}
As discussed above, there are scenarios encountered in practice where standard RAG techniques fail to retrieve contextual information for a query without introducing noise (irrelevant content)~\citep{yoran2024robustrag}. Empirical studies show that RAG efficiency is highly dependent on retriever accuracy, with ``hard negatives"---documents that are contextually similar but incorrect---leading to increased noise and compromised performance. For such cases, we strongly recommend exploring the option of continual learning (even if fine tuning via LoRA rather than full- fledged learning) of LLMs~\citep{hu2021lora, dettmers2023qlora}. The whole point of LLMs in the first place has been to make sense of the vast and diverse corpora of text and then to be able to answer a query even if the answer to the query is hidden across a sparse set of disparate text snippets across the corpora~\citep{brown2020fewshot}.
In such settings, pretraining the model to encode broader factual relationships improves recall and precision, even in the absence of reliably clean retrieval targets.

\textbf{Illustrative Example:}
Suppose your corpus contains mailing list posts, code review comments, PoCs, and advisories. Prompt-based RAG is hampered by the lack of structure and prevalence of off-topic noise. Continual pretraining (or fine-tuning) enables the LLM to internalize factual relationships, even when explicit context linking is absent.
    
\subsubsection{Robustness to noise or fake information}
The unsupervised learning of LLMs as adopted during its pre-training (or continual pre-training) is known to be relatively more robust to noise in the text (in contrast to noise in a prompt)~\citep{brown2020fewshot}. Thus, even if a corpus is noisy and contains different kinds of data such as tables, code, text, LLM learning on such corpus will be more robust in comparison to RAG. To elaborate further, suppose there is a corpus which contains legitimate as well as fake information. For RAG, it is quite possible that the fake information is retrieved as relevant information and then LLM will respond with a faulty answer to the query~\citep{huang2023hallucination_survey}. In contrast, an LLM would tend not to update its weight parameters on such fake information unless it is highly prevalent across the corpus. (Note: this statement is simplified for readers. One of the key goals of modern deep learning of models like LLMs is denoising, i.e. minimizing the signal-to-noise ratio).

\textbf{Concrete scenario:} During a high-profile vulnerability campaign, threat actors pump fake PoCs and advisories into the public web. A RAG pipeline may serve those fakes to users. But a continually (robustly) trained LLM, learning across broader context and regularization, is less susceptible to such targeted misinformation unless the signal is truly dominant.
For threat intelligence and incident response, such robustness is critical to operational safety.
    
\subsubsection{Unsupervised learning for advanced reasoning}
One of the ways to perceive unsupervised learning of LLMs is that in each iteration of the weight parameters update, the LLM is implicitly reasoning upon the input text and assessing its truth and coherence with respect to its present knowledge that is embedded in its weight parameters~\citep{wei2022chainofthought}. This reasoning happens several times since a single input text is processed multiple times during the learning process (it is not necessary to process multiple times, however we recommend so from our experience training tens of models for the cybersecurity tasks). Moreover, even different pieces of input text processed during the training are implicitly reasoned with respect to each other, thus prioritizing a piece of information for learning which aligns with the whole knowledge base (corpus). All of these factors contribute to the more robust ingestion of new knowledge by LLMs to answer a question. 
    
As one may imagine, this is highly relevant for the cybersecurity domain, since it is going to be a common phenomenon to have more and more falsified information about vulnerabilities in the future that would be easily crafted by AI agents deployed by attackers~\citep{shahriar2025agentic_security}. Of course, LLM training on its own may not be sufficient to achieve robustness to such fake information. More robust training strategies and agentic approaches are warranted for AI-driven cybersecurity. 

\textbf{Key insight:} LLMs' capacity for cross-document, multi-step reasoning means that misleading or incomplete information in a single document is downweighted when better-correlated evidence appears elsewhere in the training set.

\begin{figure}[t]
  \centering
\includegraphics[width=0.8\linewidth]{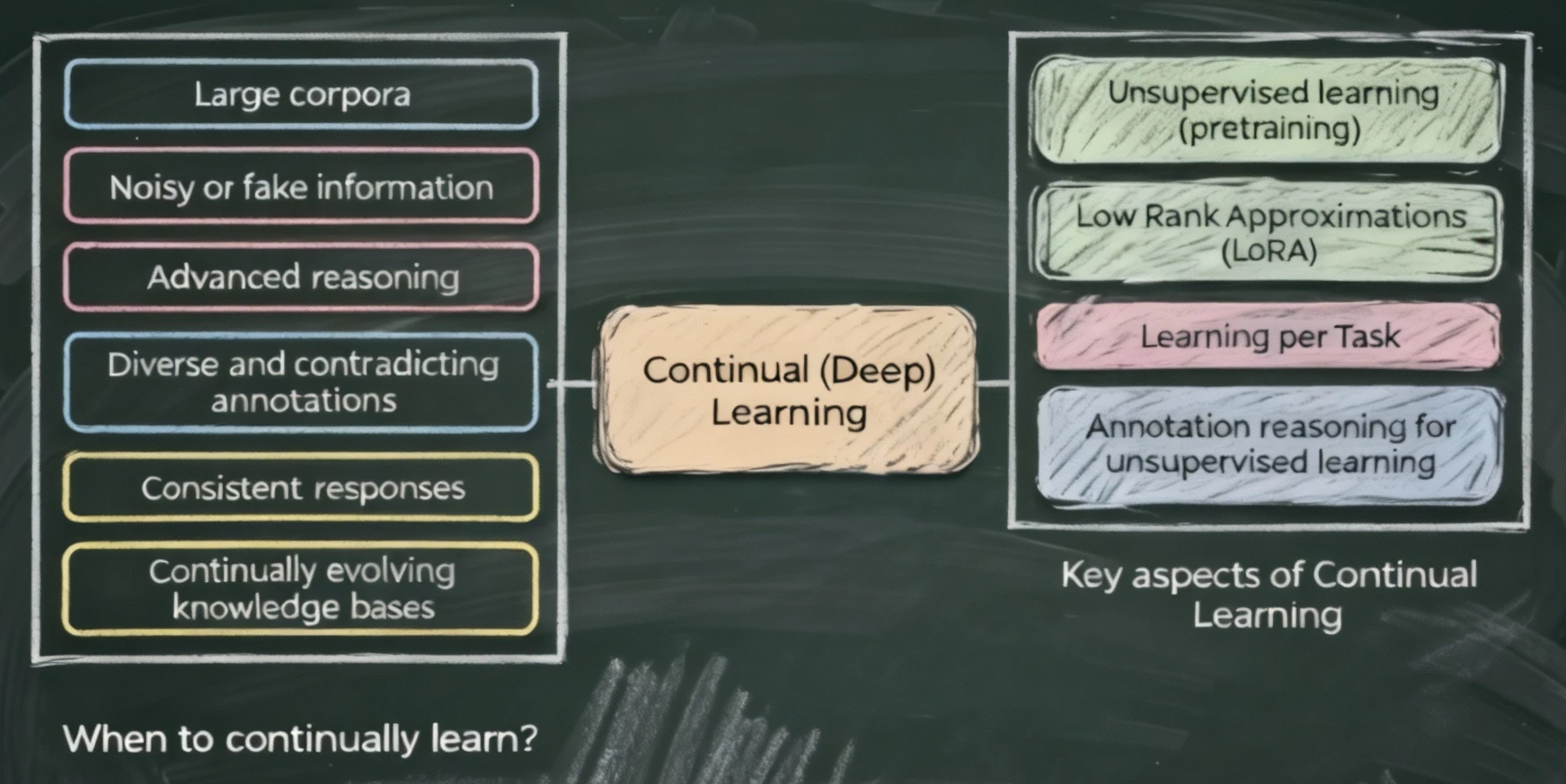}
  \caption{This figure summarizes the practical guidance on adapting large language models (LLMs) for real-world cybersecurity. The left side highlights scenarios when to continual learn LLMs: dealing with very large or messy data sources, filtering out noise or misinformation, supporting (implicit or explicit) advanced reasoning, resolving conflicting or inconsistent annotations, maintaining consistent answers across many queries, and keeping pace with fast-changing knowledge bases. The right side presents best practices for continual learning we suggest from our experience: using unsupervised learning (same objective as used for pretraining of LLMs) to absorb new information efficiently, adopting lightweight fine-tuning methods such as LoRA, tailoring learning for specific tasks instead of relying on a single general-purpose model, and focusing on training LLMs to understand the reasoning behind expert annotations—not just the labels themselves.
}
  \label{fig:continual_learning}
\end{figure}
    
\subsubsection{Large Datasets with diverse and often contradicting annotations}
For some tasks, labeled datasets are available in large sizes (on the order of tens of thousands of examples). For instance, different security advisories (such as NVD, Redhat, Red Hat, Bitnami, GHSA, etc.) provide their severity assessment for a vulnerability including the assessment for each of the factors in a CVSS vector such as Attack Vector, Attack Complexity, etc.~\citep{first2024cvss4}. Having augmented such labeled datasets with textual rationales of the labels (via agents), one can train an LLM so as to encapsulate the individual domain knowledge, bias, and intelligence of each advisory into the weight parameters of the model. This is particularly important given that conflicting CVSS scores are the norm rather than the exception~\citep{vulncheck2023cvss_accuracy}. 
    
Considering the scale of such large datasets, RAG alone, without any training, would not be an optimal use of knowledge bases. Moreover, when there are label disagreements between annotators (for instance, different CVSS vectors from different vendors), it is a natural choice to train an LLM on such datasets since the conflicted pieces of information would be implicitly reasoned for their merit with respect to each other and the present knowledge of the LLM (embedded in its weight parameters).

In scenario such as vulnerability triage, continual fine-tuning on annotated corpora with diverse, sometimes contradictory opinions improves model ability to synthesize consensus rationales or highlight disagreements for human analysts---enhancing trust.

\textbf{Example:}
When multiple vendors assign conflicting CVSS scores, continual learning enables the LLM to spot implicit heuristics (e.g., "Vendor A always considers internal attacks less severe than Vendor B") and adjust answers accordingly, rather than just parroting all available scores.
    
\subsubsection{Consistent responses across queries}
Imagine having one thousand queries to process for a given task. It is quite possible that an LLM would be stochastic in its responses, reasoning differently even for similar queries, leading to inconsistent outputs and interpretations~\citep{huang2023hallucination_survey}. As an example, a triage agent can decide to downgrade the severity assessment for a vulnerability issue due to lack of direct Internet exposure 80\% most of the time while ignoring it 20\% of the time. This kind of inconsistency can undermine trust in automated systems. On the other hand, if the responses to all the queries are used as a training set for fine tuning the LLM, the model would learn from its own responses as if it is a student as well as a teacher~\citep{huang2023selfimprove}. 
    
To understand it further, imagine a scenario where a human examiner receives answer sheets of 1000 students for evaluation. Supposedly, the examiner has high reasoning skills but not fully conversant with the subject of examination. In such case, if the examiner reads all the answer sheets before starting to evaluate any of those, the examiner can learn (gain knowledge) from the answer sheets. In essence, the examiner can act as the student while the students serve as a teacher collectively. Naturally, evaluation of the answer sheets by the examiner after having reviewed all the sheets would be more consistent and accurate in contrast to a direct evaluation. This is applicable even though answer sheets are noisy with some of the answers being completely wrong. The basic underlying assumption is that the examiner has the intelligence to differentiate wrong from right especially given the joint access to all the answers. 
    
Same principle applies to LLMs. By fine tuning an LLM on its own responses for the queries, the LLM can be enabled to generate more consistent and accurate responses in the next iteration. We have leveraged this concept across various problems including mitigation of vulnerabilities.
Iterative self-tuning, as in~\citep{huang2023selfimprove}, can systematically reduce randomness and improve repeatability of AI-driven triage at enterprise scale.

\subsubsection{Continually evolving knowledge bases}

An obvious scenario that warrants continual pretraining of LLMs is when the knowledge bases continually evolve, as we routinely witness in the domain of cybersecurity~\citep{ibrahim2024continual_pretrain, yildiz2024continual_insights}. 
    
Although it is technically possible to optimize the RAG pipelines per dynamic nature of the knowledge bases, it becomes an involved engineering task often reliant upon heuristics for its efficiency. This undermines the value of LLMs in the first place, i.e. serving as a single unit of all the knowledge and intelligence.
On the other hand, continual pretraining of LLMs on such new knowledge is less engineering and more science~\citep{ke2023continualpretrain}. Though it requires more specialized skills to establish the pipelines for continual pretraining, it is simpler and more cost-efficient to manage such pipelines in the long term, and it centralizes adaptation logic in the model rather than scattering it across many bespoke retrieval components.

Automating continual adaptation as new advisories, exploits, and mitigations land daily or weekly preserves model currency and prevents decision drift---crucial for enterprises operating at scale.
            
\begin{keypointbox}[When to continually learn LLMs?]
\begin{itemize}
    \item \textbf{Large, noisy, heterogeneous corpora}: mailing lists, code reviews, PoCs, advisories, and logs all mixed together.
    \item \textbf{Adversarial or fake information}: attackers seed fake PoCs or advisories; training tends to be more robust than per-query retrieval.
    \item \textbf{Contradictory annotations at scale}: thousands of CVEs with conflicting CVSS scores or vendor assessments.
    \item \textbf{Need for consistency over many queries}: batch decisions (e.g., org-wide triage) require models that behave consistently across similar cases.
    \item \textbf{Rapidly evolving knowledge bases}: you are ingesting new vulnerabilities, techniques, and mitigations daily or weekly.
    \item In all such cases, keeping the model itself up to date via continual (unsupervised) pretraining is often more scalable and less brittle than endlessly tuning RAG heuristics.
\end{itemize}
\end{keypointbox}

\section{Conclusion}

Precision cybersecurity is an arms race—one where static rules and static knowledge bases are quickly outpaced by both attackers and new defensive measures. 

The interplay of modern GenAI techniques—the raw recall of RAG~\citep{lewis2020rag, gao2024rag_survey}, the structure of knowledge graphs~\citep{pan2024graphrag}, the adaptive muscle of continual learning~\citep{shi2024continualsurvey}—makes it possible to forge robust, auditable, and dynamic pipelines for cybersecurity insight and automation.

The lessons from AI's evolution in fields like healthcare now echo in our domain: structure, openness, and constant, evidence-based refinement beat one-size-fits-all solutions. To win in cybersecurity, system designers must reach for every tool—embedding similarity~\citep{reimers2019sbert}, knowledge graphs~\citep{edge2024graphrag_microsoft}, continual learning~\citep{ibrahim2024continual_pretrain}—and combine them creatively and pragmatically.

In sum: Avoid monolithic approaches. Leverage structure wherever possible. Treat continual learning as a first-class citizen, not a last-resort patch. And always, always audit your pipelines for drift and adversarial manipulation.
    
We look forward to a new generation of open, reproducible, and interpretable AI-powered cybersecurity tools that make defenders nimbler and our digital world safer.

In the future, we shall share insights on the other topics relevant for AI in cybersecurity including but not limited to: (probabilistic) judge LLMs, guardrails for preserving privacy and ensuring privileged access, online learning of LLMs, knowledge distillation, reinforcement learning, domain specific model compression, learning for (implicit and explicit) reasoning, adversarial learning. 
    

            


\bibliography{references}

\bibliographystyle{plainnat}

\end{document}